\documentclass[reprint,amsmath,amssymb,aps,prl,superscriptaddress]{revtex4-2} 
\usepackage[textsize=small]{todonotes}
\usepackage{graphicx}
\usepackage[hidelinks]{hyperref}

\makeatletter
\let\old@makecaption=\@makecaption
\usepackage{subcaption}
\let\@makecaption=\old@makecaption
\makeatother

\begin{document}

\preprint{APS/123-QED}

\title{How Topology Shapes the Phase Behavior of Polyelectrolytes}

\author{David Beyer}
\email{david.beyer@mpinat.mpg.de}
\thanks{These authors contributed equally to this work.}
\affiliation{Institute for Computational Physics, University of Stuttgart, D-70569 Stuttgart, Germany}
\affiliation{Present Address: Department of Theoretical and
Computational Biophysics, Max Planck Institute for Multidisciplinary Sciences, D-37077 Göttingen, Germany}
\author{Pierre J. Walker}%
\email{pjwalker@caltech.edu}
\thanks{These authors contributed equally to this work.}
\affiliation{Division of Chemistry and Chemical Engineering, California Institute of Technology, Pasadena, California 91125, USA}
\author{Lena Tarrach}%
\affiliation{Institute for Computational Physics, University of Stuttgart, D-70569 Stuttgart, Germany}
\author{Zhen-Gang Wang}%
\affiliation{Division of Chemistry and Chemical Engineering, California Institute of Technology, Pasadena, California 91125, USA}
\author{Christian Holm}%
\affiliation{Institute for Computational Physics, University of Stuttgart, D-70569 Stuttgart, Germany}

\date{\today}

\begin{abstract}
We develop a topology-specific theory of polyelectrolyte coacervation using the random phase approximation and apply it to both simple and complex coacervation.
Our results for stars and dendrimers show that more compact chain topologies display a greater propensity for liquid-liquid phase separation, as a function of both Bjerrum length and salt concentration.
For mixtures of different topologies, we demonstrate that differences in polymer topology alone are sufficient to drive multiphase coacervation of polyelectrolytes, which we rationalize in terms of an effective $\chi$ parameter.
Analysis of a simplified global phase diagram reveals that the propensity for such topology-driven phase separation is largest at a finite molecular weight.
Overall, our results establish polymer topology as a powerful design lever for tuning the phase diagram of charged macromolecules independently of molecular weight, net charge, and monomer chemistry, since changes in topology enable fine-tuning of the effective charge density without altering these molecular characteristics.
\end{abstract}

\maketitle

\begin{figure*}[ht]
\includegraphics[width=0.85\linewidth]{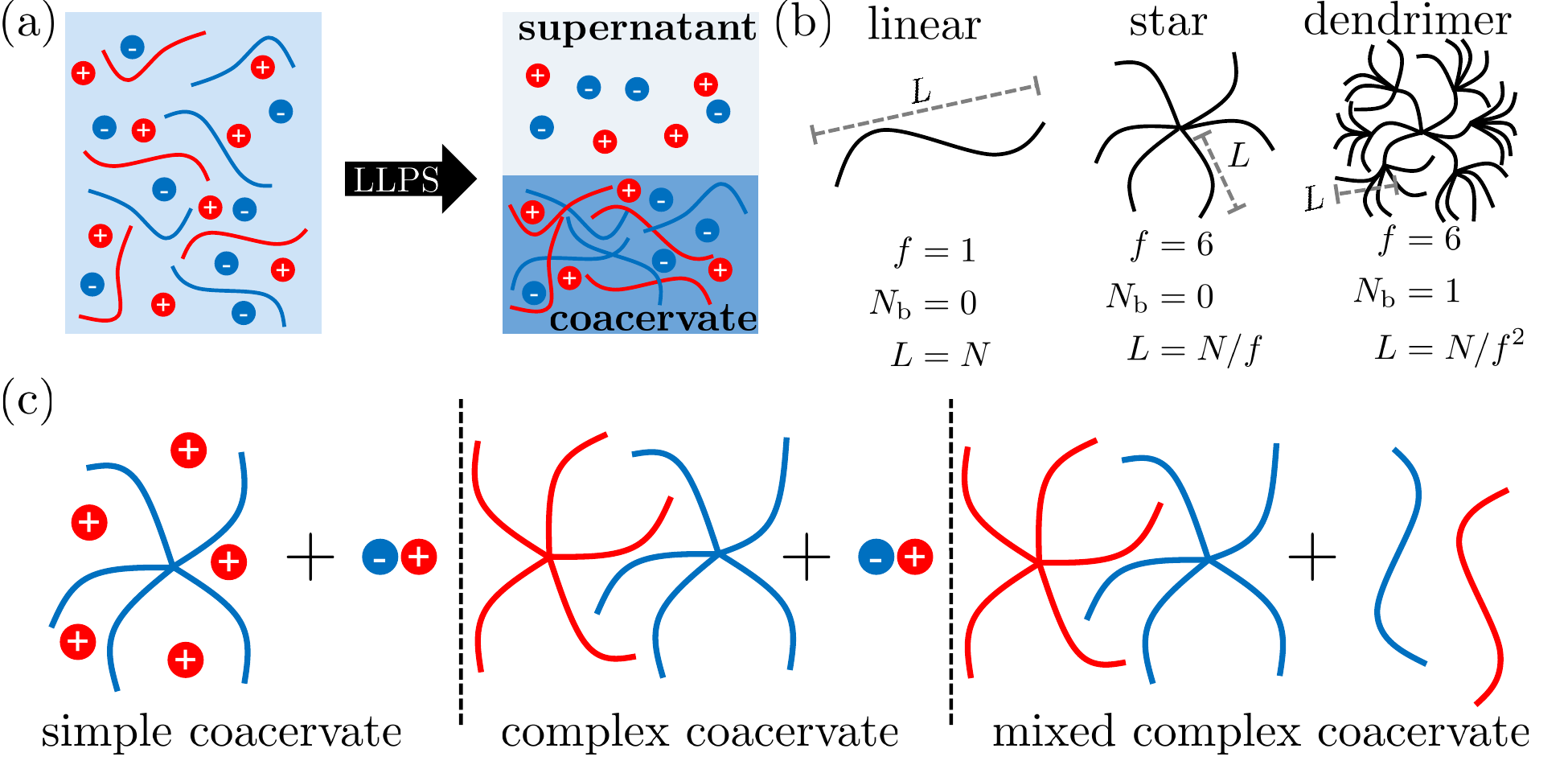}
\caption{
\label{fig:schematic}
(a): Polyelectrolyte solutions can split into a supernatant with a low polyelectrolyte concentration and a coacervate with a high polyelectrolyte concentration through LLPS.
Blue and red chains correspond to polyanions and polycations, respectively.
(b): Schematic representation of the different polymer topologies investigated in this work: linear chains, stars, and dendrimers. 
Note that for visual clarity, the different topologies are not shown at the same scale.
(c): Overview of the different mixtures of polyelectrolytes studied in this work: simple coacervates (left panel), complex coacervates (middle panel), and salt-free, mixed complex coacervates (right panel).
}
\end{figure*}

{\em Introduction.} 
The phase behavior of charged macromolecules plays an important role in cell biology \cite{hymanLiquidLiquidPhaseSeparation2014, bananiBiomolecularCondensatesOrganizers2017}, soft matter science \cite{overbeekPhaseSeparationPolyelectrolyte1957, e.singRecentProgressScience2020, rumyantsevPolyelectrolyteComplexCoacervates2021, muthukumarPhysicsChargedMacromolecules2023}, and various technological applications \cite{ramirezmarreroFundamentalsDesignRules2026}.
For example, charged biopolymers can spontaneously phase separate, leading to droplet-like structures known as biomolecular condensates \cite{bananiBiomolecularCondensatesOrganizers2017, choiPhysicalPrinciplesUnderlying2020, galvanettoExtremeDynamicsBiomolecular2023, daiUnlockingElectrochemicalFunctions2024}.
An analogous phenomenon occurs for synthetic polyelectrolytes \cite{overbeekPhaseSeparationPolyelectrolyte1957, e.singRecentProgressScience2020, rumyantsevPolyelectrolyteComplexCoacervates2021}: as illustrated in \autoref{fig:schematic} (a), aqueous solutions of polyelectrolytes can split into a dense coacervate phase and a dilute supernatant through liquid-liquid phase separation (LLPS).
Depending on the composition of the system---either solely consisting of polyelectrolytes with the same sign of charge or containing both polyanions and polycations---the coacervate is called a simple coacervate or a complex coacervate.
Recent works have revealed that mixtures of different polyelectrolytes may even split into multiple distinct coacervate phases \cite{luMultiphaseComplexCoacervate2020, chenMultiphaseCoacervatesDriven2021, jacobsTheorySimulationMultiphase2023, wangInterfacialTensionsPolyelectrolyte2024, agrawalChargeDensityMismatch2025, spruijtChapter6Multiphase2023, luControllingMultiphaseCoacervate2025}---so-called ``multiphase coacervation.''

A key question in polyelectrolyte physics is identifying the parameters that govern the coacervate phase behavior.
Previous studies have demonstrated that LLPS can be controlled by adjusting parameters such as temperature \cite{ylitaloElectrostaticCorrelationsTemperatureDependent2021}, salt concentration \cite{spruijtBinodalCompositionsPolyelectrolyte2010, liPhaseBehaviorSalt2018}, and pH \cite{weinbreckComplexCoacervationWhey2003, ausserwogerBiomolecularCondensatesSustain2026}.
Furthermore, the phase behavior of charged macromolecules is also influenced by properties intrinsic to polymers, including charge density and net charge \cite{neitzelPolyelectrolyteComplexCoacervation2021, chenComplexationOppositelyCharged2022, luoTheoryCondensateSize2025}, chain stiffness \cite{shakyaRoleChainFlexibility2020}, and architectural parameters such as the sequence of charged residues \cite{linSequenceSpecificPolyampholytePhase2016, pakSequenceDeterminantsIntracellular2016, linTheoriesSequenceDependentPhase2018, changSequenceEntropybasedControl2017, danielsenMolecularDesignSelfcoacervation2019, lytleDesigningElectrostaticInteractions2019, yuComplexCoacervationStatistical2021, adachiPredictingHeteropolymerInteractions2024}.
Another architectural parameter that could influence the phase behavior of polyelectrolytes is the polymer topology.
However, although there exist a small number of publications studying the coacervation of polyelectrolytes with non-linear topologies \cite{qinCriticalityConnectivityMacromolecular2016, johnstonEffectCombArchitecture2017, lytleTuningChainInteraction2018, tagliabueCanOppositelyCharged2021, stanoElectrostaticallyCrossLinkedReversible2021, yuStructureDynamicsHybrid2023, 
chenMultiphaseCoacervationPolyelectrolytes2023,
zhangSelfAssemblyStarPolyelectrolytesVarious2024, stevensImpactLightlyBranched2024, gurelCharacterizingStructuralConformation2025}, the majority of studies focus on linear chains, presumably because of their prominence in biological systems and ease of synthesis. 
Previous works studying the effects of molecular architecture on polyelectrolyte coacervation invariably varied both chain topology and other parameters such as molecular weight simultaneously \cite{qinCriticalityConnectivityMacromolecular2016, lytleTuningChainInteraction2018},  making it difficult to isolate the effects of chain topology.
Moreover, we are not aware of any systematic theoretical investigation establishing the phase behavior of complex mixtures containing polyelectrolytes of different topologies.
To address these open questions, we develop a minimal, topology-specific theory of polyelectrolyte coacervates and apply it to simple and complex coacervation, revealing several topology-driven effects, including topology-driven multiphase coacervation.

{\em Theory.} 
Because the space of polymer topologies is astronomically large, we use a class of simple model architectures that allow us to systematically vary the chain topology via a small number of easily interpretable parameters.
The model architectures studied in the following are dendrimers, which have previously been studied, for example, in the context of binary polyelectrolyte complexation \cite{welchDendrimerPolyelectrolyteComplexationModel2000}.
We consider dendrimers with $f$ arms emerging from the central core (see \autoref{fig:schematic} (b)).
Each arm branches into $f-1$ branches (these branches then branch into $f-1$ sub-branches, etc); in total, there are $N_{\mathrm{b}}$ generations of branchings.
For a given topology, the number of monomers between two branchings is always taken to be the same ($L$), resulting in three parameters that fully characterize our model topologies ($\mathcal{T}=(f,N_{\mathrm{b}},N)$): the number of arms ($f$), the number of generations ($N_{\mathrm{b}}$), and the overall molecular weight ($N$).
To focus solely on the effect of topology, we set the degree of polymerization of the polymers to a constant value, $N=200$, in the following.
We represent the polymer chains using a continuous Gaussian chain model, which allows us to derive a simple closed-form expression for the single-chain structure factors and is justified given that $L\gg 1$ for all model architectures considered.

To construct phase diagrams of coacervates consisting of polyelectrolyte chains with different topologies, we need a topology-specific expression for the free energy. 
Approaches such as the Voorn-Overbeek theory \cite{overbeekPhaseSeparationPolyelectrolyte1957} or liquid-state theory \cite{mullerSimulationHardTriatomic1993,zhangSaltingOutSaltingInPolyelectrolyte2016,zhangSaltPartitioningComplex2018,zhangPolyelectrolyteComplexCoacervation2018,ylitaloElectrostaticCorrelationsTemperatureDependent2021,marshallHigherOrderClassical2011,marshallThreeNewBranched2013,zhangModelingThermodynamicProperties2018} cannot account for differences in polymer topology.
Therefore, we combine the Flory-Huggins free energy with a Coulomb correlation term derived within the random phase approximation (RPA) \cite{fredricksonEquilibriumTheoryInhomogeneous2005}, which has been widely used to study the phase behavior of polyelectrolytes \cite{borueStatisticalTheoryWeakly1988, 
mahdiPhaseDiagramsSaltFree2000, leeComplexCoacervationField2008a, 
qinCriticalityConnectivityMacromolecular2016,
linRandomphaseapproximationTheorySequencedependent2017, 
delaneyTheoryPolyelectrolyteComplexation2017,
rumyantsevControllingComplexCoacervation2019, 
chenMultiphaseCoacervatesDriven2021}.
The free energy, $F$, for a mixture of salt, polyelectrolytes of various topologies, and neutral solvent reads
\begin{widetext}
\begin{align}
    \frac{\beta F b^3}{V} = \sum_i \frac{\phi_i}{N_i}\ln\phi_i +  \frac{1}{2}\sum_{i,j}\chi_{ij}\phi_i\phi_j+\frac{b^3}{4\pi^2}\int_{0}^{\infty} \mathrm{d}k\,k^2\left[\ln\left(1+\frac{4\pi l_\mathrm{B}}{k^2b^3}\sum_i\phi_i\sigma_i^2g_i(k;\mathcal{T}_i)\right)-\frac{4\pi l_\mathrm{B}}{k^2b^3}\sum_i\phi_i\sigma_i^2\right],
\label{eq:free-energy}
\end{align}
\end{widetext}
with sums running over all species $i$.
The first two terms are the Flory-Huggins free energy, where $\phi_i$ and $N_i$ are the volume fraction and molecular weight of species $i$, respectively. 
$b$ is the Kuhn length, which, for simplicity, is assumed to be the same for all polymers.
Furthermore, we set the volume of small ions and solvent molecules to $b^3$.  
$\chi_{ij}$ is the interaction parameter of species $i$ and $j$, which is set to $\chi_{ij}=1/2$ for polymer-solvent interactions to emulate a $\theta$-solvent and set to zero for all other interactions \cite{rumyantsevControllingComplexCoacervation2019}.
The final term accounts for electrostatic correlations of polyelectrolytes and small ions at the RPA level.
Here, $k$ is the wave number and $l_\mathrm{B}\equiv e^2/4\pi\epsilon_0\epsilon_\mathrm{r}k_\mathrm{B}T$ is the Bjerrum length of the system, with the elementary charge $e$, the vacuum permittivity $\epsilon_0$, and the dielectric constant of the solvent $\epsilon_\mathrm{r}$.
$\sigma_i$ is the charge fraction of species $i$, i.e. the average fraction of elementary charges per monomer; we set $\sigma_i = 0.2$ for all polyelectrolyte species and $\sigma_i=1$ for salt ions.
$g_i(k; \mathcal{T}_i)$ is the single-chain structure factor of species $i$, which explicitly depends on the corresponding topology $\mathcal{T}_i$ \footnote{See Supplemental Material for details}.
To find the composition of coexisting phases, we have to equate their osmotic pressures as well as the electrochemical potentials of the ionic species, subject to an incompressibility constraint $\sum_i \phi_i =1$.
In practice, we use an implementation of \autoref{eq:free-energy} in the Clapeyron.jl software package \cite{walkerClapeyronjlExtensibleOpenSource2022}, which uses automatic differentiation to evaluate derivatives of the free energy and allows for efficient numerical computation of phase diagrams \footnote{All code used for producing the results shown in this work are provided in the Supplemental Material.}.

\begin{figure}[ht!]
\includegraphics[width=\columnwidth]{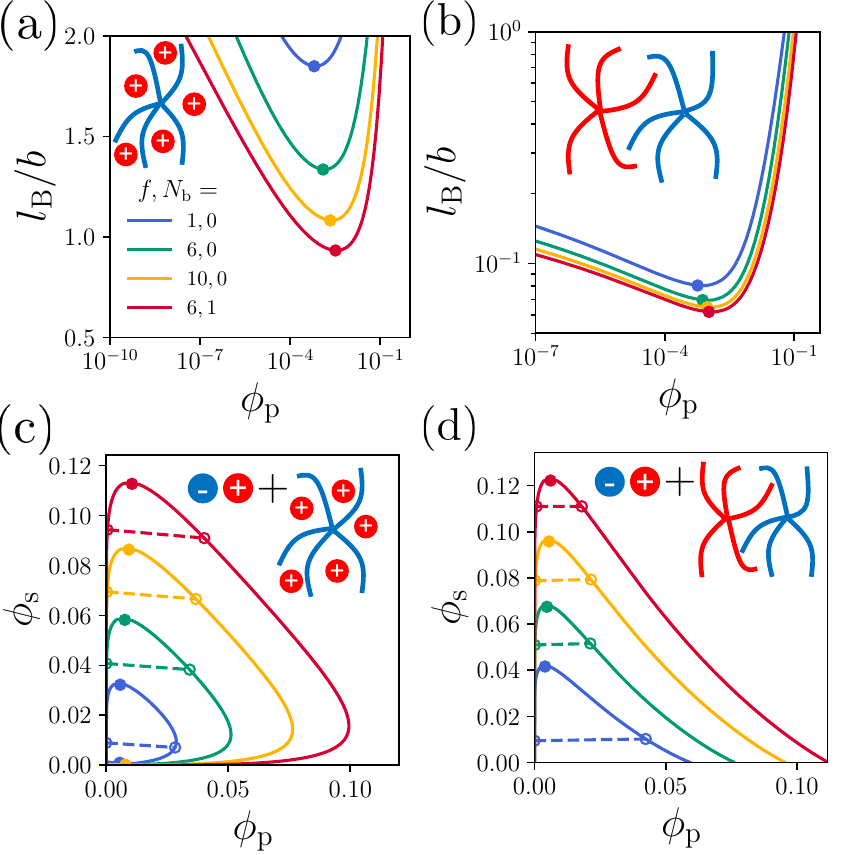}
\caption{
\label{fig:polyelectrolyte_salt}
Phase diagrams of simple (subfigure (a) and (c)) and complex (subfigure (b) and (d)) coacervates of varying topology.
Subfigures (a) and (b) show salt-free phase diagrams as a function of the polymer volume fraction and the Bjerrum length.
Subfigures (c) and (d) show diagrams in the presence of salt as a function of the polymer volume fraction and the salt volume fraction (the Bjerrum length is set to $l_{\mathrm{B}} = b$). 
Solid lines correspond to the co-existence boundary; dashed lines correspond to tie lines connecting coexisting phases (empty markers); solid markers denote the critical points.
}
\end{figure}

{\em Results.} 
We now apply our theory to model systems of increasing complexity (\autoref{fig:schematic} (c)).
First, we consider a salt-free simple coacervate, i.e., a polyelectrolyte with counterions in the absence of salt.
\autoref{fig:polyelectrolyte_salt} (a) shows that this system undergoes LLPS for a sufficiently high Bjerrum length (``simple coacervation'').
As the Bjerrum length is reduced, the electrostatic interactions, and thus the driving force for LLPS, weaken, causing the coacervate and supernatant branches to approach each other and finally merge at the critical point.
Despite the fact that the total charge and molecular weight of each topology shown in the plot are identical, a comparison shows that the critical Bjerrum length decreases in the order $(1,0)\rightarrow (6,0)\rightarrow (10,0)\rightarrow (6,1)$ (using the notation introduced in the legend of \autoref{fig:polyelectrolyte_salt} (a)).
This increased tendency to phase separate is driven solely by changes in topology and stems from the more compact nature of branched architectures, which leads to a locally increased charge density and, therefore, to stronger electrostatic correlations driving LLPS. 
The increased compactness is evident from the RPA structure factors of the different topologies (see ESI Figure S2), which can in turn be connected to an effective charge density (see ESI for details).
Notably, topology-induced changes in charge density are distinct from those reported for linear polyelectrolytes \cite{chenMultiphaseCoacervatesDriven2021}, where changes in linear charge density are invariably accompanied by variations in monomer size or polymer chemical composition. 
In contrast, changes in topology allow the monomer properties and charge density to be decoupled, enabling fine-tuning of the phase behavior while maintaining a fixed polyelectrolyte chemistry and molecular weight.
For salt-free complex coacervates, we observe a largely analogous behavior (\autoref{fig:polyelectrolyte_salt} (b)).
However, the critical Bjerrum length for complex coacervates is more than an order of magnitude smaller than for simple coacervates.
This higher propensity for LLPS is caused by strong electrostatic correlations between the polyelectrolytes driving the phase split, as opposed to weaker polyelectrolyte-counterion correlations in simple coacervates.
\begin{figure*}[ht!]
\includegraphics[width=0.95\linewidth]{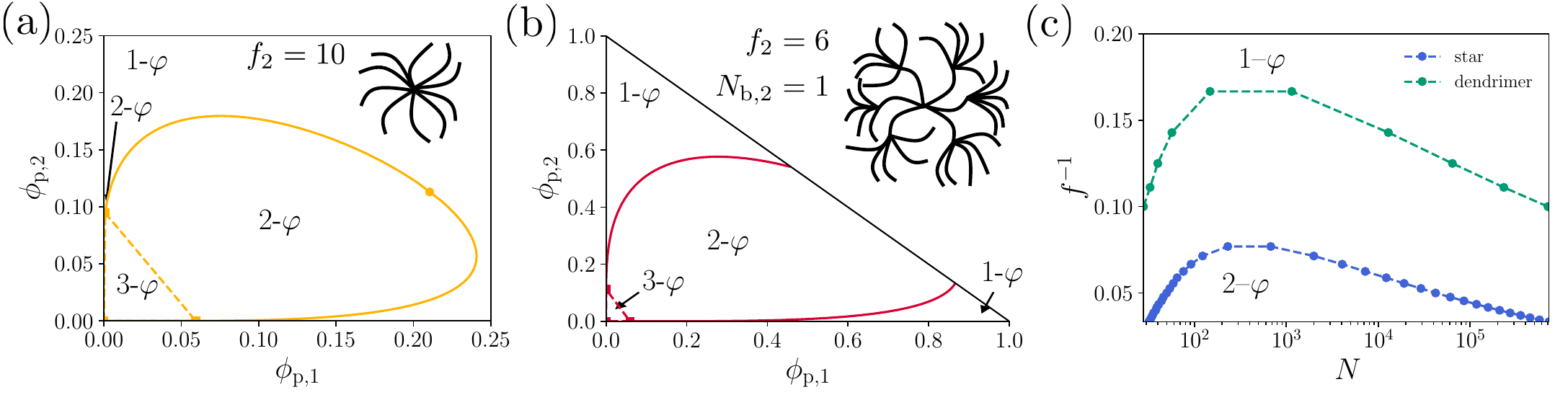}
\caption{
\label{fig:mix}
Phase behavior of mixed complex coacervates where symmetric pairs are considered for all topologies, i.e. the same number of polyanions and polycations is always added to the system for a particular topology.
``1-$\varphi$'', ``2-$\varphi$'' and ``3-$\varphi$'' denote regions where one, two or three phases coexist, respectively.
(a): Phase diagram of a salt-free mixture of linear chains (volume fraction $\phi_{\mathrm{p},1}$) and stars ($f_2=10$, volume fraction $\phi_{\mathrm{p},2}$).
(b): Phase diagram of a salt-free mixture of linear chains (volume fraction $\phi_{\mathrm{p},1}$) and dendrimers with one generation of branching ($f_2=6$, $N_{\mathrm{b},2}=1$, volume fraction $\phi_{\mathrm{p},2}$).
(c): Global phase diagram of a salt- and solvent-free blend of linear chains with stars (blue) and dendrimers ($N_{\mathrm{b}}=1$, green) with varying degree of polymerization ($N$) and degree of asymmetry (taken as $f^{-1}$). 
Points represent conditions at which the blend of polyelectrolytes will not phase separate at $l_\mathrm{B}=b$. 
If a mixture lies below these curves, the addition of solvent will eventually result in three-phase separation.  
}
\end{figure*}

In the next step, we fix the Bjerrum length to $l_{\mathrm{B}} = b$ and study coacervation in the presence of added salt \footnote{For many polymers, $b\approx 0.4-0.6\,\mathrm{nm}$ \cite{giannottiInterrogationSingleSynthetic2007}, meaning that this Bjerrum length would be slightly smaller than that of water.}.
As shown in \autoref{fig:polyelectrolyte_salt} (c), for simple coacervation, all topologies exhibit the same qualitative phase behavior. 
The most prominent feature of the phase diagrams is a reentrant coacervation: at low salt concentration, an increase in salt leads to a broadening of the two-phase region, i.e., the polymer concentration in the coacervate increases. 
For star topologies, where the salt-free critical $l_\mathrm{B}$ is greater than $b$, the addition of salt even induces phase separation, as indicated by the critical points at low salt concentrations.
The observed reentrant behavior occurs due to initial increases in electrostatic correlations upon the addition of salt, strengthening the driving force for coacervation. 
Conversely, at sufficiently high salt concentration, electrostatic screening begins to dominate, weakening the driving force for coacervation and causing the two phase region to shrink and eventually close at the critical point.
Changes in topology lead to a trend consistent with the salt-free case: more compact architectures exhibit a larger two-phase region and thus higher salt resistance, with the phase separation propensity increasing in exactly the same order as for the salt-free case.
At the same time, the critical polymer concentration slightly increases, which also mirrors the behavior observed for the salt-free systems. 
Consistent with other theories on simple coacervates \cite{zhangSaltingOutSaltingInPolyelectrolyte2016,zhangPolyelectrolyteComplexCoacervation2018}, negatively-sloped tie lines are observed due to differences in electrostatic screening between the supernatant and the coacervate (for a further discussion in ESI).
For complex coacervates (\autoref{fig:polyelectrolyte_salt} (d)), the phase behavior is largely similar.
However, we do not observe reentrant coacervation, i.e., the two-phase region monotonically shrinks with increasing salt concentration.
This behavior originates from the stronger electrostatic correlations in complex coacervates: because the correlations between polyelectrolyte chains are already quite strong in the absence of salt, adding salt has only a minor effect that is overwhelmed by the increase in electrostatic screening.
Moreover, we observe that in contrast to the simple coacervates and prior works, the tie lines are positively-sloped, due to the stronger electrostatic correlations in the complex coacervate (further details are provided in the ESI).

Having studied simple and complex coacervates in which only a single polymer topology is present in the system, we now turn to mixtures of different topologies. 
To avoid unnecessary complexity, we consider complex coacervates containing pairs of linear polyanions and polycations, mixed with pairs of polyanions and polycations with a non-linear topology (see \autoref{fig:schematic} (c)), and set $l_{\mathrm{B}} = b$.
For systems with minor topological asymmetries, e.g., linear chains and stars with $f=6$ arms, almost ideal mixing occurs within the coacervate phase and the observed phase behavior is largely analogous to the results reported in the previous sections.
In contrast, for systems with stronger asymmetries in polymer topology, such as linear chains mixed with 10-arm stars (\autoref{fig:mix} (a)), a more complex phase diagram emerges: in addition to single-phase behavior (``1-$\varphi$'') and two-phase coexistence (``2-$\varphi$''), we also observe three-phase coexistence (``3-$\varphi$'').
These three coexisting phases, corresponding to the vertices of the dashed triangle, are a supernatant with a low overall polymer concentration, a coacervate phase in which the linear chains are enriched, and \emph{another} coacervate phase with a high concentration of stars.
Notably, this three-phase coexistence only occurs in a regime where the concentrations of both polymer topologies are fairly low.

The driving force behind the observed three-phase separation (and, at higher concentrations, the two-phase separation between polyelectrolytes of different topologies) ultimately stems from differences in chain topology:
the effect of topology on the electrostatic interactions between chains manifests itself as a change in the effective charge densities of the polyelectrolytes and, in fact, the whole single-chain structure factors.
Building on the work of Chen et al. \cite{chenMultiphaseCoacervatesDriven2021, chenMultiphaseCoacervationPolyelectrolytes2023}, in the ESI we show that the thermodynamic compatibility of polyelectrolyte species $i$ and $j$ with the same charge fraction, $\sigma$, but different topologies, $\mathcal{T}_i$ and $\mathcal{T}_j$, is governed by an effective $\chi$ parameter:
\begin{align}
    \chi_{ij}^{\mathrm{eff}} \equiv \chi_{ij} + \frac{\sigma^4}{2}\int_{0}^{\infty} \mathrm{d}k\,\alpha(k)\left[g_i(k;\mathcal{T}_i)-g_j(k;\mathcal{T}_j)\right]^2.
\end{align}
Here, $\alpha(k)$ is a function that depends on the electrostatic coupling strength, as well as the topologies and concentrations of all species present in the system. 
If the difference between the topologies $\mathcal{T}_i$ and $\mathcal{T}_j$ and thus the difference between the single-chain structure factors is sufficiently large, $\chi_{ij}^{\mathrm{eff}}$ can exceed its critical value, resulting in topology-driven phase separation.
The observed closure of the two-phase separation between topologies at higher concentrations is driven primarily by charge screening, which weakens electrostatic correlations in both phases.
Mathematically, this manifests itself as a decrease in $\alpha(k)$, which in turn leads to a lower value of $\chi_{ij}^{\mathrm{eff}}$.
Overall, our analytical result for $\chi_{ij}^{\mathrm{eff}}$ establishes topology as a key determinant of LLPS: topology modifies intramolecular charge-correlation spectra, and electrostatic
fluctuations convert differences between these spectra into an effective thermodynamic incompatibility, which can drive phase separation.

For a mixture of linear chains with dendrimers (\autoref{fig:mix} (b)), a slightly different phase behavior is observed: first, the concentration regime in which three-phase coexistence occurs is slightly larger.
Second, the two-phase region no longer terminates at a critical point and instead intersects the solvent-free edge $\phi_{\mathrm{p},2}=1-\phi_{\mathrm{p},1}$.
As in \autoref{fig:mix} (a), the driving force of multiphase coacervation is the difference in the intramolecular charge correlations, which are described by single-chain structure factors.
Due to weak electrostatic screening, the topology-driven LLPS is particularly pronounced at low polyelectrolyte concentrations. 
However, the differences in topology are so large that even in the solvent-free case, screening is insufficient to close the two-phase region.
On the whole, the mixture of linear chains and dendrimers therefore exhibits a much larger concentration range where some form of topology-driven phase separation---either into two or three phases---occurs.

To characterize the effect of topological asymmetries on the global phase behavior, we consider a solvent- and salt-free blend comprised of linear polyelectrolytes and polyelectrolytes of arbitrary topology. 
For a given molecular weight, $N$, we then calculate the minimal degree of asymmetry (characterized by $f$) needed to drive LLPS between the different topologies.
As shown in \autoref{fig:mix} (c), both for stars and dendrimers ($N_{\mathrm{b}}=1$), we observe a non-monotonic dependence of the asymmetry $f$ required to drive LLPS on the molecular weight $N$. 
Because dendrimers have an innately greater degree of topological asymmetry than stars, a lower degree of branching is needed for topology-induced LLPS, resulting in a larger two-phase region.
The observed non-monotonicity can be understood as follows: on the left branch, where the molecular weight is low, electrostatic correlations are weak.
Therefore, to induce LLPS, the charge density must be increased, resulting in a positive correlation between $N$ and $f^{-1}$.
Conversely, for large $N$, the RPA expression for the electrostatic correlation free energy decays as $N^{-3/2}$, which is faster than the entropy of mixing, $N^{-1}$ (see ESI for more details).
In order to induce LLPS, there must thus be a sufficiently high degree of topological asymmetry to overcome the entropy of mixing, resulting in a decrease of $f^{-1}$ with $N$.
These results are consistent with the scaling behavior of $\chi_{ij}^\mathrm{eff}-\chi_{ij}$ that we derive in the ESI: for small $N$ we have $\chi_{ij}^\mathrm{eff}-\chi_{ij}\sim N^{5/2}$, and for large $N$ we have $\chi_{ij}^\mathrm{eff}-\chi_{ij}\sim N^{-3/2}$.
Taken together, our analytical results therefore suggest that topology-induced LLPS should be most pronounced for polyelectrolytes of finite molecular weight, which is in agreement with the numerically obtained phase diagrams.

{\em Conclusion.} 
In summary, we have theoretically studied the influence of polymer topology on the phase behavior of polyelectrolyte coacervates.
The emerging picture is that more compact chain topologies show a greater propensity for LLPS, as a function of both Bjerrum length and salt concentration.
Furthermore, we also investigated the phase behavior that results when polyelectrolytes of the same molecular weight but different topologies are mixed.
These systems exhibited a more complex phase behavior, including three-phase coexistence driven solely by differences in topology, which we could rationalize in terms of an effective $\chi$ parameter.
Analysis of a simplified global phase diagram revealed that the propensity for such topology-driven LLPS has a non-monotonic functional dependence on the molecular weight, becoming increasingly suppressed for very small or large molecular weights.
Overall, our results establish polymer topology as a powerful design lever for tuning the phase diagram of charged macromolecules independently of molecular weight, net charge, and monomer chemistry, since changes in topology enable fine-tuning of the effective charge density without altering these molecular characteristics.
Looking ahead, it will be interesting to test our predictions using simulations that do not rely on the approximations of our RPA theory and ultimately through experiments.
In particular, it would be worthwhile to investigate topology-modulated quantities and phenomena that are not accessible within our theory, including single-chain conformations, material properties, and polymer dynamics.

\begin{acknowledgments}
D.B. acknowledges the support of the German-American Fulbright Program.
D.B. and C.H. acknowledge the German Research Foundation (DFG) for funding within the Research Unit FOR2811 ``Adaptive Polymer Gels with Model Network Structure'' under grant 423791428 along with grant 397384169 (TP7). 
Z.-G.W. acknowledges funding from the Hong Kong Quantum AI Lab, AIR@InnoHK of the Hong Kong Government. 
C.H. acknowledges funding through the DFG grants 451980436 and 268449726, as well as EXC 2075-390740016.
The authors thank Prof. Dr. Alejandro Gallegos and Dr. Pablo M. Blanco for their detailed comments on the manuscript.
\end{acknowledgments}

\bibliography{topology_coacervates}

\end{document}


\title{Supplemental Material: How Topology Shapes the Phase Behavior of Polyelectrolytes}

\author{David Beyer}
\email{david.beyer@mpinat.mpg.de}
\thanks{These authors contributed equally to this work.}
\affiliation{Institute for Computational Physics, University of Stuttgart, D-70569 Stuttgart, Germany}
\affiliation{Present Address: Department of Theoretical and
Computational Biophysics, Max Planck Institute for Multidisciplinary Sciences, D-37077 Göttingen, Germany}
\author{Pierre J. Walker}%
\email{pjwalker@caltech.edu}
\thanks{These authors contributed equally to this work.}
\affiliation{Division of Chemistry and Chemical Engineering, California Institute of Technology, Pasadena, California 91125, USA}
\author{Lena Tarrach}%
\affiliation{Institute for Computational Physics, University of Stuttgart, D-70569 Stuttgart, Germany}
\author{Zhen-Gang Wang}%
\affiliation{Division of Chemistry and Chemical Engineering, California Institute of Technology, Pasadena, California 91125, USA}
\author{Christian Holm}%
\affiliation{Institute for Computational Physics, University of Stuttgart, D-70569 Stuttgart, Germany}

\maketitle

\section{Details on the Theory}
\subsection{RPA Expression for the Free Energy}
The total free energy ($F$) for a system comprised of polyelectrolytes of various topologies, simple salts, and neutral solvent can be obtained as:

\begin{align}
    \frac{\beta F b^3}{V} = \sum_i \frac{\phi_i}{N_i}\ln\phi_i +  \frac{1}{2}\sum_{i,j}\chi_{ij}\phi_i\phi_j+\frac{b^3}{4\pi^2}\int_{0}^{\infty} \mathrm{d}k\,k^2\bigg[\ln\left(1+\frac{4\pi l_\mathrm{B}}{k^2b^3}\sum_i\phi_i\sigma_i^2g_i(k;\mathcal{T}_i)\right)-\frac{4\pi l_\mathrm{B}}{k^2b^3}\sum_i\phi_i\sigma_i^2\bigg].
\label{eq:free-energy}
\end{align}
In this equation, $\phi_i$ is the volume fraction of species $i$, $N_i$ is the degree of polymerization of species $i$, $b$ is the Kuhn length of the polymers (which is taken to be equal equal to the size of salt ions and solvent molecules), and $\chi_{ij}$ is the interaction parameter of species $i$ and $j$.
$l_\mathrm{B}\equiv e^2/4\pi\epsilon_0\epsilon_\mathrm{r}k_\mathrm{B}T$ is the Bjerrum length of the system, with the elementary charge $e$, vacuum permittivity $\epsilon_0$, and dielectric constant of the solvent $\epsilon_\mathrm{r}$.
$\sigma_i$ is the charge fraction of species $i$, and $g_i(k; \mathcal{T}_i)$ is the single-chain structure factor of species $i$. 
For salt ions, we have $g_i=1$.
For a continuous chain with a dendrimer-like topology, characterized by a degree of polymerization $N$, branching number $f$, and $N_\mathrm{b}$ generations, the single-chain structure factor is defined as
\begin{equation}
g(k;\mathcal{T})
=
\frac{1}{N}
\int_0^N \mathrm{d}s
\int_0^N \mathrm{d}s'
\left\langle
e^{i\mathbf{k}\cdot[\mathbf{r}(s)-\mathbf{r}(s')]}
\right\rangle,
\end{equation}
where $s$ and $s'$ denote contour positions along the polymer backbone. Assuming Gaussian chain statistics, the characteristic function for monomers belonging to the same edge takes the form
\begin{equation}
\left\langle
e^{i\mathbf{k}\cdot[\mathbf{r}(s)-\mathbf{r}(s')]}
\right\rangle
=
\exp\!\left(
-\frac{k^2b^2|s-s'|}{6}
\right).
\end{equation}
The correlations between points belonging to the same edge of the dendrimer, therefore, generate the standard Debye function $D(x)$,
\begin{equation}
\frac{1}{L^2}
\int_0^L \mathrm{d}s
\int_0^L \mathrm{d}s'
\exp\!\left(
-\frac{k^2b^2|s-s'|}{6}
\right) = \frac{2}{x^2}
\left(
e^{-x}
+x
-1
\right) = D(x),
\end{equation}
where we introduced the abbreviation $x\equiv Lk^2b^2/6$ and $L$, which is the length of a single graph edge.
Correlations between a branch point and the positions on an adjacent edge generate the auxiliary function $h(x)$,
\begin{equation}
\frac{1}{L}
\int_0^L \mathrm{d}s
\exp\!\left(
-\frac{k^2b^2 s}{6}
\right) = \frac{1}{x}
\left(
1-e^{-x}
\right) = h(x).
\end{equation}
The full dendrimer structure factor is obtained by combining same-edge and inter-edge correlations. 
The latter depend only on the graph distance between edges and can be collected into a generating function $G(z;f,N_\mathrm{b})$, where $z=e^{-x}$. 
The resulting single-chain structure factor is
\begin{align}
g(k;\mathcal{T})=\frac{N}{E(f,N_\mathrm{b})}\bigg[
D(x)
+
\frac{G(z;f,N_\mathrm{b})}{E(f,N_\mathrm{b})}
h^2(x)\bigg],
\end{align}
where
\begin{align}
    E = f\frac{(f-1)^{N_{\mathrm{b}}+1}-1}{f-2}
\end{align}
is the total number of edges in the dendrimer, allowing us to calculate the length of a single edge,
\begin{equation}
    L = \frac{N}{E(f,N_\mathrm{b})}.
\end{equation}
For $N_\mathrm{b}=0$, we have:
\begin{equation}
    G(z)=f(f-1),
\end{equation}
and for $N_\mathrm{b}=1$, we have:
\begin{equation}
    G(z) = (f-1)f(f+1)+2f(f-1)^2z+f(f-1)^3z^2.
\end{equation}

\subsection{Large-$N$ Asymptotics of the RPA Free Energy}
In the main text, we discuss the global phase diagram of mixtures of two different topologies.
To understand the large-$N$ behavior of topology-driven phase separation, let us consider a four-component system containing pairs of oppositely charged polyelectrolytes with topology $\mathcal{T}_A$ and topology $\mathcal{T}_B$, all with the same degree of polymerization $N$ and charge fraction $\sigma$. 
The RPA contribution to the free energy is
\begin{align}
    &\frac{\beta F_\mathrm{RPA} b^3}{V}
    =
    \frac{b^3}{4\pi^2}\int_{0}^{\infty} \mathrm{d}k\,k^2
    \bigg[\ln\left(
    1+\frac{4\pi l_\mathrm{B}\sigma^2}{k^2b^3}
    \left[
    \phi_A g_A(k;\mathcal{T}_A)
    +\phi_B g_B(k;\mathcal{T}_B)
    \right]
    \right) -
    \frac{4\pi l_\mathrm{B}\sigma^2}{k^2b^3}
    \left[
    \phi_A 
    +\phi_B
    \right]
    \bigg].
\end{align}
The electrostatic contribution which governs topology-driven phase separation is the excess RPA contribution relative to the two pure-topology reference systems, which for a solvent-free system is given by
\begin{align}
\begin{split}
    \frac{\beta \Delta F_\mathrm{RPA} b^3}{V}
    =
    \frac{b^3}{4\pi^2}
    \int_{0}^{\infty} \mathrm{d}k\,k^2
    \bigg[&\ln\left(
    1+\frac{4\pi l_\mathrm{B}\sigma^2}{k^2b^3}
    \left[
    \phi_A g_A(k;\mathcal{T}_A)
    +(1-\phi_A)g_B(k;\mathcal{T}_B)
    \right]
    \right) \\
    &-\phi_A
    \ln\left(
    1+\frac{4\pi l_\mathrm{B}\sigma^2}{k^2b^3}
    g_A(k;\mathcal{T}_A)
    \right) -(1-\phi_A)
    \ln\left(
    1+\frac{4\pi l_\mathrm{B}\sigma^2}{k^2b^3}
    g_B(k;\mathcal{T}_B)
    \right)
    \bigg].
\end{split}
\label{eq:delta_f}
\end{align}
We now introduce the simplified notation
\begin{align}
\zeta=\frac{4\pi l_\mathrm{B}\sigma^2}{b^3},
\qquad
g_i(k;\mathcal{T}_i)=N D_i(k^2N),
\end{align}
where $D_i$ is the generalized Debye function of species $i$.
Note that for economy of notation, we have suppressed the dependence of $D_i$ on the other architectural parameters, since here we are only interested in the scaling with $N$.
We also define the composition-weighted, generalized Debye function
\begin{align}
\bar D(k^2N,\phi_A)
=
\phi_A D_A(k^2N)
+
(1-\phi_A)D_B(k^2N).
\end{align}
Splitting the first logarithm in \autoref{eq:delta_f} into $\phi_A$ and $(1-\phi_A)$ contributions gives
\begin{align}
\frac{\beta \Delta F_\mathrm{RPA} b^3}{V}
    =
    \frac{b^3}{4\pi^2}
    \int_{0}^{\infty} \mathrm{d}k\,k^2
    \bigg[\phi_A
    \ln\left(
    \frac{
    1+\frac{N\zeta}{k^2}\bar D(k^2N,\phi_A)
    }{
    1+\frac{N\zeta}{k^2}D_A(k^2N)
    }
    \right) +(1-\phi_A)
    \ln\left(
    \frac{
    1+\frac{N\zeta}{k^2}\bar D(k^2N,\phi_A)
    }{
    1+\frac{N\zeta}{k^2}D_B(k^2N)
    }
    \right)
    \bigg].
\end{align}
To extract the large-$N$ dependence, we rescale the integration variable according to
\begin{align}
q^2=k^2N,
\qquad
k^2\,\mathrm{d}k = N^{-3/2}q^2\,\mathrm{d}q.
\end{align}
This gives
\begin{align}
    &\frac{\beta \Delta F_\mathrm{RPA} b^3}{V}
    =
    \frac{b^3}{4\pi^2N^{3/2}}
    \int_{0}^{\infty} \mathrm{d}q\,q^2
    \bigg[\phi_A
    \ln\left(
    \frac{
    1+\frac{N^2\zeta}{q^2}\bar D(q^2,\phi_A)
    }{
    1+\frac{N^2\zeta}{q^2}D_A(q^2)
    }
    \right) +(1-\phi_A)
    \ln\left(
    \frac{
    1+\frac{N^2\zeta}{q^2}\bar D(q^2,\phi_A)
    }{
    1+\frac{N^2\zeta}{q^2}D_B(q^2)
    }
    \right)
    \bigg].
\end{align}
In the large-$N$ limit, the integrand admits an asymptotic expansion in powers of $N^{-2}$, with a leading order term that is independent of $N$.
Asymptotically, the excess RPA contribution thus becomes
\begin{align}
    \frac{\beta \Delta F_\mathrm{RPA} b^3}{V}
    \approx 
    \frac{b^3}{4\pi^2N^{3/2}}
    \int_{0}^{\infty} \mathrm{d}q\,q^2
    \bigg[\phi_A
    \ln\left(
    \frac{\bar D(q^2,\phi_A)}{D_A(q^2)}
    \right) +(1-\phi_A)
    \ln\left(
    \frac{\bar D(q^2,\phi_A)}{D_B(q^2)}
    \right)
    \bigg],
\end{align}
which scales as
\begin{align}
\frac{\beta \Delta F_\mathrm{RPA} b^3}{V}
\sim N^{-3/2}.
\end{align}
This scaling is physically intuitive: in the melt phase, as $N$ becomes sufficiently large, the influence of branching diminishes because monomers are increasingly ``unable'' to distinguish whether neighboring monomers belong to the same branch or to different branches.

\subsection{Effective $\chi$ Parameter}
To elucidate the mechanism of topology-driven phase separation, we use an approach previously established by Chen et al. \cite{chenMultiphaseCoacervatesDriven2021, chenMultiphaseCoacervationPolyelectrolytes2023}.
Let us consider the free energy density of a general mixture,
\begin{widetext}
\begin{align}
    \mathfrak{f} \equiv \frac{\beta F b^3}{V} = \underbrace{\sum_i \frac{\phi_i}{N_i}\ln\phi_i +  \frac{1}{2}\sum_{i,j}\chi_{ij}\phi_i\phi_j}_{\mathfrak{f}_{\mathrm{F-H}}}+\underbrace{\frac{b^3}{4\pi^2}\int_{0}^{\infty} \mathrm{d}k\,k^2\left[\ln\left(1+\frac{4\pi l_\mathrm{B}}{k^2b^3}\sum_i\phi_i\sigma_i^2g_i(k;\mathcal{T}_i)\right)-\frac{4\pi l_\mathrm{B}}{k^2b^3}\sum_i\phi_i\sigma_i^2\right]}_{\mathfrak{f}_{\mathrm{RPA}}}.
\label{eq:free-energy-chi}
\end{align}
\end{widetext}
The thermodynamic stability of a bulk phase is governed by the curvature of this free energy density.
The second-order variation of $\mathfrak{f}$ under infinitesimal density variations $\delta \phi_i$ can be written as 
\begin{align}
    \delta ^2\mathfrak{f} = \sum_i \frac{1}{N_i\phi_i}\delta \phi_i^2 + \frac{1}{2}\sum_{i,j}\left[\chi_{ij}+\frac{\partial^2 \mathfrak{f}_{\mathrm{RPA}}}{\partial\phi_i\partial \phi_j}\right]\delta\phi_i\delta\phi_j.
\end{align}
The Hessian matrix of the RPA contribution reads
\begin{align}
    \frac{\partial^2 \mathfrak{f}_{\mathrm{RPA}}}{\partial\phi_i\partial \phi_j} &= -\frac{b^3}{4\pi^2}\int_{0}^{\infty} \mathrm{d}k\,k^2\left(\frac{4\pi l_\mathrm{B}}{k^2b^3}\right)^2\frac{\sigma_i^2\sigma_j^2g_i(k;\mathcal{T}_i)g_j(k;\mathcal{T}_j)}{\left(1+\frac{4\pi l_\mathrm{B}}{k^2b^3}\sum_l\phi_l\sigma_l^2g_l(k;\mathcal{T}_l)\right)^2}\\
    &= -\int_{0}^{\infty} \mathrm{d}k\,\alpha(k)\sigma_i^2\sigma_j^2g_i(k;\mathcal{T}_i)g_j(k;\mathcal{T}_j),
\end{align}
where we defined the abbreviation
\begin{align}
    \alpha(k)\equiv \frac{b^3k^2}{4\pi^2}\left(\frac{4\pi l_\mathrm{B}}{k^2b^3}\right)^2\frac{1}{\left(1+\frac{4\pi l_\mathrm{B}}{k^2b^3}\sum_l\phi_l\sigma_l^2g_l(k;\mathcal{T}_l)\right)^2}.
\end{align}
Following the approach of Chen et al. \cite{chenMultiphaseCoacervatesDriven2021, chenMultiphaseCoacervationPolyelectrolytes2023}, we can now rewrite the variation $\delta^2 \mathfrak{f}$ using the incompressibility condition $\sum_i \delta \phi_i = 0$:
\begin{align}
\begin{split}
    \delta ^2\mathfrak{f} =& \sum_i \frac{1}{N_i\phi_i}\delta \phi_i^2 + \frac{1}{2}\sum_{i,j}\left[\chi_{ij}+\frac{\partial^2 \mathfrak{f}_{\mathrm{RPA}}}{\partial\phi_i\partial \phi_j}\right]\delta\phi_i\delta\phi_j\\
    =&\sum_i \frac{1}{N_i\phi_i}\delta \phi_i^2 + \frac{1}{2}\sum_{i,j}\left[\chi_{ij}+\frac{\partial^2 \mathfrak{f}_{\mathrm{RPA}}}{\partial\phi_i\partial \phi_j}+A_i+A_j\right]\delta\phi_i\delta\phi_j.
\end{split}
\end{align}
Note that this equation holds for \emph{arbitrary} $A_i,A_j\in\mathbb{R}$.
Setting 
\begin{align}
    A_i = \frac{1}{2}\int_{0}^{\infty} \mathrm{d}k\,\alpha(k)\sigma_i^4g_i^2(k;\mathcal{T}_i),
\end{align}
$\delta^2 \mathfrak{f}$ can be written as 
\begin{align}
\begin{split}
    \delta ^2\mathfrak{f} =& \sum_i \frac{1}{N_i\phi_i}\delta \phi_i^2 + \frac{1}{2}\sum_{i,j}\chi_{ij}^{\mathrm{eff}}\delta\phi_i\delta\phi_j,
\end{split}
\end{align}
with the effective $\chi$ parameter,
\begin{align}
    \chi_{ij}^{\mathrm{eff}} \equiv \chi_{ij} + \frac{1}{2}\int_{0}^{\infty} \mathrm{d}k\,\alpha(k)\left[\sigma_i^2g_i(k;\mathcal{T}_i)-\sigma_j^2g_j(k;\mathcal{T}_j)\right]^2.
\end{align}
Let us now consider two polyelectrolyte species $i$ and $j$ that have the same charge densities, $\sigma_i=\sigma_j=\sigma$, but different topologies, $\mathcal{T}_i\neq\mathcal{T}_j$.
In this case, the effective interaction between the species is governed by
\begin{align}
    \chi_{ij}^{\mathrm{eff}} \equiv \chi_{ij} + \frac{\sigma^4}{2}\int_{0}^{\infty} \mathrm{d}k\,\alpha(k)\left[g_i(k;\mathcal{T}_i)-g_j(k;\mathcal{T}_j)\right]^2.
\end{align}
Therefore, if the two topologies are sufficiently different, such that $\chi_{ij}^{\mathrm{eff}}$ exceeds a critical value, differences in topology alone can drive liquid-liquid phase separation.
To examine the $N$-dependence of $\chi_{ij}^{\mathrm{eff}}$, we consider a salt-free mixture of polyelectrolytes with the same molecular weight and perform the same transformations as in the previous section.
We find that $\alpha(k)$ exhibits the following $N$-dependence:
\begin{equation}
    \alpha(k)\equiv \frac{b^3q^2}{4\pi^2N}\left(\frac{4\pi l_\mathrm{B}N}{q^2b^3}\right)^2\frac{1}{\left(1+\frac{4\pi l_\mathrm{B}N}{q^2b^3}\sum_l\phi_l\sigma_l^2g_l(q;\mathcal{T}_l)\right)^2}.
\end{equation}
Therefore, in the limit of large $N$, we obtain the scaling behavior
\begin{equation}
    \alpha(k)\sim \frac{1}{N^3}\rightarrow \chi_{ij}^{\mathrm{eff}}-\chi_{ij}\sim N^{-3/2}.
\end{equation}
Conversely, for small $N$ we get
\begin{equation}
    \alpha(k)\sim N \rightarrow \chi_{ij}^{\mathrm{eff}}-\chi_{ij}\sim N^{5/2}
\end{equation}

\FloatBarrier
\section{Additional Results}
\subsection{Salt Partitioning}
\begin{figure}[t]
    \centering
    \includegraphics[width=1\linewidth]{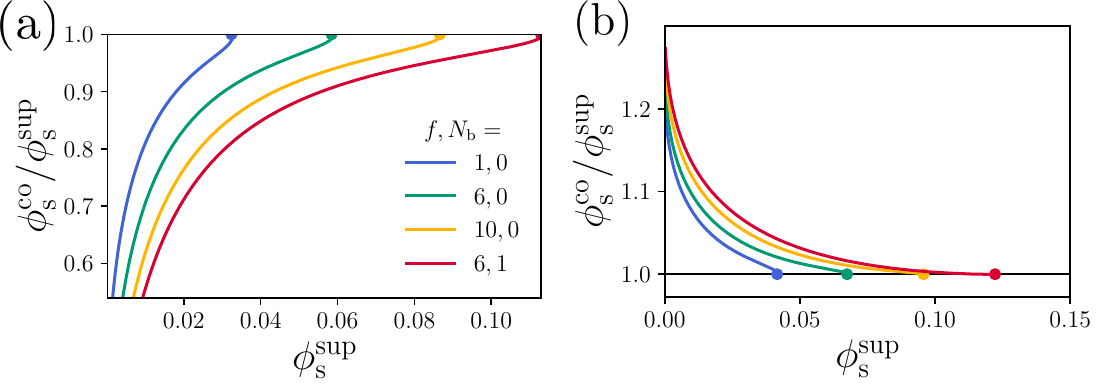}
    \caption{Salt partitioning in a simple (a) and complex (b) coacervate as a function of salt concentration in the supernatant with varying polyelectrolyte topology. For all systems, $l_\mathrm{B}=b$.}
    \label{fig:salt_partitioning}
\end{figure}
While a few tie-lines are included within the added-salt figures in the main text, it is often of interest to examine how the partitioning of salt varies across the phase diagram. In \autoref{fig:salt_partitioning}, we present the partition coefficient of salts within a simple (a) and complex (b) coacervate as a function of salt concentration in the supernatant with varying polyelectrolyte topology.
\autoref{fig:salt_partitioning} (a) demonstrates that for a simple coacervate, the partition coefficient is consistently smaller than unity; i.e., salt is rejected by the simple coacervate and the salt concentrations in the two phases only equalize at the critical point.
Comparing different topologies, we observe that more compact molecular architectures reject salt more strongly, with the ability to reject salt increasing in exactly the same order as the phase separation propensity. This type of behavior is expected in simple coacervates, as the higher ion concentration in the supernatant results in greater electrostatic screening in this phase. 
Thus, with effectively weaker electrostatic interactions, the salt will preferentially partition into the supernatant.
For the complex coacervate, we observe that ions preferentially partition into the coacervate rather than the supernatant (\autoref{fig:salt_partitioning} (b)), which continues up to the critical point. 
The preferential partitioning into the coacervate is due to the stronger electrostatic correlations within this phase due to the presence of the oppositely charged polyelectrolytes, in contrast to the simple coacervate system containing a single polyelectrolyte with its small counterion.

\subsection{Structure Factors and Effective Charge Density}
\begin{figure}[t]
    \centering
    \includegraphics[width=0.5\linewidth]{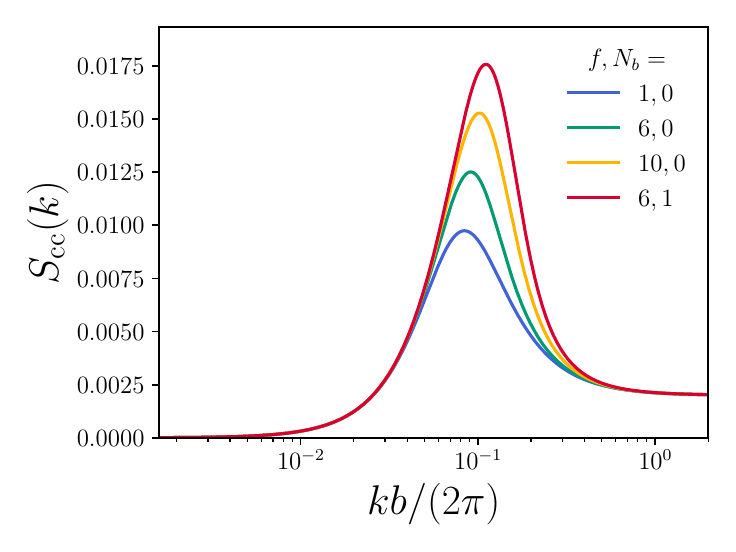}
    \caption{Charge--charge structure factor of a simple polyelectrolyte of varying topology with its counterion at $l_\mathrm{B}=b$, $\sigma=0.2$, $\phi_\mathrm{p}=0.01$ and $N=200$.}
    \label{fig:sf}
\end{figure}
Within the random phase approximation, it is possible to obtain the charge--charge structure factor as:
\begin{equation}
    S_\mathrm{cc}(k)=\frac{\sum_i \phi_i\sigma_i^2g_i(k;\mathcal{T}_i)}{1+\frac{4\pi l_\mathrm{B}}{k^2b^3}\sum_i \phi_i\sigma_i^2g_i(k;\mathcal{T}_i)}\,.
\end{equation}
The resulting structure factor for a dilute simple polyelectrolyte with its counterion is shown in \autoref{fig:sf}. 
In all cases, a maximum appears at a finite $k$, implying compact conformations. 
These maximum becomes more pronounced and shifts to higher $k$ as the polymer topology becomes more branched, implying further compactification of the single chain.
This behavior demonstrates the higher local charge density of branched topologies. 

To derive asymptotic scaling laws for the effective charge density, we  examine single-chain charge--charge structure factor:
\begin{equation}
    S_\mathrm{p}(k)=\frac{\sigma_\mathrm{p}^2g_\mathrm{p}(k;\mathcal{T}_\mathrm{p})}{1+\frac{4\pi l_\mathrm{B}}{k^2b^3}\sum_i \phi_i\sigma_i^2g_i(k;\mathcal{T}_i)}.
\end{equation}
At infinite dilution, $\phi_\mathrm{p}\rightarrow 0$, keeping the ion concentration fixed, $\phi_i>0$, we have:
\begin{equation}
    S_\mathrm{p}(k)=\sigma_\mathrm{p}^2g_\mathrm{p}(k;\mathcal{T}_\mathrm{p})\frac{1}{1+\frac{4\pi l_\mathrm{B}}{k^2b^3}\sum_i \phi_i\sigma_i^2}\,.
\end{equation}
Recognizing the second term in the denominator as the square of the inverse Debye screening length, $\kappa_\mathrm{s}^2 =\frac{4\pi l_\mathrm{B}}{b^3}\sum_i\phi_i\sigma_i^2$, we can write:
\begin{equation}
    S_\mathrm{p}(k)=\sigma_\mathrm{p}^2g_\mathrm{p}(k;\mathcal{T}_\mathrm{p})\frac{k^2}{k^2+\kappa_\mathrm{s}^2}\,.
\end{equation}
\begin{equation}
    \frac{S_\mathrm{p}(k)}{\sigma_\mathrm{p}L}=\left(\frac{2}{(Lb^2k^2/6)^2}
\left[
e^{-(Lb^2k^2/6)}
+(Lb^2k^2/6)
-1
\right]
+\frac{G(z;f,N_\mathrm{b})}{E(f,N_\mathrm{b})}\left[\frac{1}{(Lb^2k^2/6)}
\left(
1-e^{-(Lb^2k^2/6)}
\right)\right]^2\right)\frac{k^2}{k^2+\kappa_\mathrm{s}^2}\,.
\end{equation}
Using $x=Lb^2k^2/6$ and defining $B(x)\equiv G(e^{-x};f,N_\mathrm{b})/E(f,N_\mathrm{b})$ and $\mathcal{K}\equiv Lb^2\kappa_\mathrm{s}^2/6$, we get
\begin{equation}
    \frac{S_\mathrm{p}(x)}{\sigma_\mathrm{p}L}=\frac{F(x)}{x(x+\mathcal{K})},
\end{equation}
where
\begin{equation}
    F(x) = 2
\left(
e^{-x}
+x
-1
\right)
+B(z)
\left(
1-e^{-x}
\right)^2.
\end{equation}
Differentiating w.r.t. $x$ to find the maximum results in
\begin{equation}
    \frac{d}{dx}\left( \frac{S_\mathrm{p}(x)}{\sigma_\mathrm{p}L}\right)=
\frac{
x(x+
\mathcal{K})F'(x)
-
(2x+\mathcal{K})
F(x)
}{
x^2(x+\mathcal{K})^2
},\label{eq:derivative_S}
\end{equation}
with
\begin{equation}
    F'(x) = 2(1-e^{-x})(1+B(x)e^{-x})+B'(x)(1-e^{-x})^2
\end{equation}
The root of equation \ref{eq:derivative_S} (which is the maximum of $S_\mathrm{p}(k)$) corresponds to the solution of the equation:
\begin{equation}
x(x+
\mathcal{K})F'(x)
-
(2x+\mathcal{K})
F(x)=0\,.
\end{equation}
At low salt concentration or large chain length, $\mathcal{K}\rightarrow 0$, the asymptotic behavior for small $\mathcal{K}$ can be obtained by expanding the equation in powers of $x$:
\begin{equation}
    (1+B_0)\mathcal{K}x^2-
    \left(\frac{1}{3}+B_0+B_1\right)x^4+O(\mathcal{K}x^3,x^5)=0.\label{eq:approx_S}
\end{equation}
For a star topology we have
\begin{equation}
    B_0=f-1,\, B_1=0,
\end{equation}
and for a dendrimer topology we have
\begin{equation}
    B_0 = f^2-1,\, B_1 = 2(f-1)^2\,.
\end{equation}
Solving equation \ref{eq:approx_S} gives the solution:
\begin{equation}
    x^*\approx \sqrt{\frac{3(1+B_0)}{1+3(B_0+B_1)}}\sqrt{\mathcal{K}}\,.
\end{equation}
In the case of a linear chain we thus obtain
\begin{equation}
    (k^*)^2\approx 3\sqrt{2}\frac{\kappa_\mathrm{s}}{N^{1/2}b}\,.
\end{equation}
In the case of a star polymer, we get
\begin{equation}
    (k^*)^2\approx 3\sqrt{2}\frac{\kappa_\mathrm{s}}{N^{1/2}b}\sqrt{\frac{f^2}{3f-2}}\,.
\end{equation}
For the dendrimer topology we get
\begin{equation}
    (k^*)^2\approx 3\sqrt{2}\frac{\kappa_\mathrm{s}}{N^{1/2}b}\frac{f^2}{3f-2}\,.
\end{equation}
Assuming that $(k^*)^3$ approximately represents the inverse volume taken up by a single chain, we can estimate the increase in charge density when going from a linear chain to a topology $\mathcal{T}$:
\begin{equation}
    \frac{\rho_\mathrm{c,eff}(\mathcal{T})}{\rho_\mathrm{c,eff}(\mathrm{chain})}\sim\frac{(k^*_\mathcal{T})^3}{(k^*_\mathrm{chain})^3}.
\end{equation}
Thus, for a star topology, we obtain
\begin{equation}
   \frac{\rho_\mathrm{c,eff}(\mathrm{star})}{\rho_\mathrm{c,eff}(\mathrm{chain})}\sim \left(\frac{f^2}{3f-2}\right)^{3/4},
\end{equation}
and for a dendrimer
\begin{equation}
   \frac{\rho_\mathrm{c,eff}(\mathrm{dendrimer})}{\rho_\mathrm{c,eff}(\mathrm{chain})}\sim \left(\frac{f^2}{3f-2}\right)^{3/2}.
\end{equation}
\bibliography{topology_coacervates}